\newcommand{\be}{\begin{equation}}\newcommand{\ee}{\end{equation}}
\newcommand{\bea}{\begin{eqnarray}}\newcommand{\eea}{\end{eqnarray}}
\newcommand{\nn}{\nonumber\\[6pt]}
\newcommand{\p}[1]{(\ref{#1})}

\newcommand{\bD}{\overline D}

\newcommand{\cU}{{\cal U}}
\newcommand{\cZ}{{\cal Z}}

\newcommand{\bV}{{\overline V}}

\newcommand{\bxi}{{\bar\xi}}
\newcommand{\bpsi}{{\bar\psi}}
\newcommand{\brho}{{\bar\rho}}
\newcommand{\bphi}{{\bar\phi}}
\newcommand{\blam}{{\bar\lambda}}
\newcommand{\bLam}{{\overline\Lambda}}

\newcommand{\wQ}{{\widetilde Q}}
\newcommand{\wS}{{\widetilde S}}
\newcommand{\wbQ}{{\widetilde{\overline Q}}}
\newcommand{\wbS}{{\widetilde{\overline S}}}

\newcommand{\bPhi}{{\overline \Phi}}

\documentclass[12pt]{article}
\usepackage{amscd,amsmath,amssymb}

\begin{document}

\thispagestyle{empty}
\vspace{2cm}
\begin{flushright}
hep-th/0407015 \\[3cm]
\end{flushright}
\begin{center}
{\Large\bf N=4 superconformal mechanics in the pp-wave limit  }
\end{center}
\vspace{1cm}

\begin{center}
{\large\bf S. Bellucci${}^{a}$,
S. Krivonos${}^{b}$, E. Orazi${}^{a,c}$ }
\end{center}

\begin{center}
${}^a$ {\it INFN-Laboratori Nazionali di Frascati,
Via E. Fermi 40, 00044 Frascati, Italy}

\vspace{0.2cm}

${}^b$ {\it Bogoliubov  Laboratory of Theoretical Physics, JINR, 141980 Dubna,
Russia}

\vspace{0.2cm}
${}^c$ {\it Dip. di Fisica, Univ. di Roma Tre, Via della Vasca Navale 84
00146 Roma, Italy} \\

\end{center}
\vspace{2cm}

\begin{abstract}
We constructed the pp-wave limit of $N=4$ superconformal mechanics with the
off-shell $({\bf 3,4,1})$ multiplet. We present the superfield and
the component actions which exhibit the interesting property that the interaction
parts are completely fixed by the symmetry.
We also explicitly demonstrate that the passing to the pp-wave limit can be achieved by
keeping at most quadratic nonlinearities in the action of (super)conformal mechanics.
\end{abstract}

\newpage
\setcounter{page}{1}
\section{Introduction}
The simplest example of the famous AdS/CFT correspondence \cite{mald1}, which relates the
string theory on $AdS_{p+2}\times S^{D-p-2}$ to extended superconformal theories in $p+1$
dimensions, is provided by the theory of a 0-brane on $AdS_2\times S^2$. Owing to the AdS/CFT
conjecture, the action describing the motion of a super 0-brane in $AdS_2\times S^2$ background
should be related to the action of $N=4$ superconformal mechanics with $SU(1,1|2)$
superconformal symmetry. The first step toward establishing this relation
was done in \cite{oldmech} where it was shown that the radial motion of a superparticle
with zero angular momentum near the horizon of an extreme Reissner-Nordstr{\"o}m black hole
is described by an $Osp(1|2)$ superconformal mechanics. A proper accounting of the angular
degrees of freedom brings us to a new variant of $N=4$ superconformal mechanics \cite{IKL1}
containing three physical bosonic fields in its supermultiplet (i.e. the radial $AdS_2$ coordinate
 and two angular coordinates parameterizing $S^2$). A Green-Schwarz-type
action for the $AdS_2\times S^2$ superparticle was constructed in \cite{gsparticle}.
After a proper gauge-fixing, the corresponding action can be related with $N=4$ superconformal
mechanics \cite{{IKN},{BGKL1}}.

Apart from $AdS_{p+2}\times S^{D-p-2}$ and flat space, there exists another maximally
supersymmetric background -- i.e. the
pp-wave background \cite{pp1}. One of the nice features of the pp-wave background is that
string theory can be solved exactly on it \cite{pp2}. Therefore, it is obviously interesting to
understand the role of AdS/CFT duality in the
Penrose limit. Usually, the pp-wave limit is considered on the "string" side
(see e.g. \cite{pp3} and references therein),
while the "conformal" side is much less understood \cite{conf1}. In this respect,
the $N=4$ superconformal
mechanics, being equivalent to a super 0-brane in $AdS_2\times S^2$ background, provides a
nice and simple toy
theory to fully understand what happens in the pp-wave limit on the "conformal" side.
Our aim in this work is to construct the
pp-wave limit of $N=4$ superconformal mechanics.

The most convenient framework for constructing superconformal quantum mechanics with extended
supersymmetries is based
on nonlinear realizations of $d=1$ superconformal groups.
\footnote{We recall that superconformal
quantum mechanics is also closely related to the integrable Calogero-Moser-type systems \cite{cal1,BGK}.} .
It was pioneered in \cite{leva} and recently
advanced in \cite{{IKL1},{IKL2},{BIKL}}. In the present paper we apply this method to consider
nonlinear realizations of the pp-wave limit of the conformal supergroup $SU(1,1|2)$. In this way
we re-derive the off-shell multiplet ({\bf 3, 4, 1}) \cite{mult} which is still useful in
the pp-wave limit and construct nontrivial off-shell superfield actions.

The paper  is organized as follows. In Section 2 we demonstrate how the nonlinear realizations
technique works
in the bosonic case by considering $AdS_2\times S^2$ conformal mechanics in the pp-wave limit.
In Section 3 we present $N=4$ superfield formulations of $N=4$  pp-wave superconformal
mechanics. The last Section is left for the summary and conclusions.

\setcounter{equation}0
\section{Conformal mechanics in the pp-wave limit}
\subsection{Conformal mechanics on $AdS_2\times S^2$}
We will start with conformal mechanics on $AdS_2\times S^2$. The symmetry underlying
this case is $so(1,2)\oplus su(2)$
\bea\label{confalg}
i\left[ D, P\right]=P,\quad i\left[ D, K\right]=-K,\quad i\left[ P, K\right]=-2D &&
\qquad so(1,2) \nn
i\left[ V_3, V\right]=-V,\quad i\left[ V_3, \bV\right]=\bV, \quad i\left[ V, \bV\right]=2 V_3
&&  \qquad su(2) \;.
\eea
We will be interested in the nonlinear realization of $SO(1,2)\times SU(2)$ in its coset space
$SO(1,2)\times SU(2)/U(1) \sim AdS_2\times S^2$. There are different choices for the
parametrization of this coset
(see e.g. \cite{IKN}) which are actually equivalent, but the simplest one is provided by
\be\label{confcoset}
g=e^{itP}e^{i\cZ K}e^{i\cU D}e^{i\Phi V + i\bPhi \bV} \;.
\ee
The coset parameters $\left\{ \cU, \cZ,\Phi,\bPhi \right\}$ are Goldstone fields depending
on the time $t$. In order
to construct an invariant action, one should calculate the left-invariant Cartan forms which
are defined in a standard way
\be\label{confform1}
g^{-1}dg=i\omega_P P+i\omega_D D+i\omega_K K+i\omega_V V+i\bar\omega_V \bV+\omega_3 V_3
\ee
and explicitly read
\bea\label{confform2}
&&\omega_P=e^{-\cU}dt,\quad \omega_D= d\cU-2 \cZ dt ,\quad \omega_K=e^{\cU}
\left( d\cZ+\cZ{}^2 dt\right) , \nn
&& \omega_V= \frac{d\Lambda}{1+\Lambda\bLam},\quad \bar\omega_V=\frac{d\bLam}{1+\Lambda\bLam},
\quad
\omega_3=\frac{\bLam d\Lambda- \Lambda d\bLam}{1+\Lambda\bLam} \;,
\eea
where
\be\label{lambda}
\Lambda = \frac {\tan \sqrt{\Phi\bPhi}}{\sqrt{\Phi\bPhi}} \Phi ,\quad
\bLam = \frac {\tan \sqrt{\Phi\bPhi}}{\sqrt{\Phi\bPhi}} \bPhi \;.
\ee
The transformation properties of the time $t$ and fields
$\left\{ \cU, \cZ, \Phi, \bPhi \right\}$ under
the group $SO(1,2)\times SU(2)$ are generated by the left action of the coset element.
For example, the
$SU(2)/U(1)$ coset transformations are generated by the left action of the element
\be
g_1=e^{ia V + i{\bar a}\bV}
\ee
and read
\be
\delta \Lambda = a +{\bar a}\Lambda^2,\quad \delta\bLam = {\bar a} + a\bLam{}^2\;.
\ee
Let us note that the Cartan forms $\omega_P,\omega_D$ and $\omega_K$ are invariant with respect
to $SO(1,2)\times SU(2)$ transformations. So we can reduce the number of independent fields
using the Inverse Higgs constraint \cite{IH}
\be\label{confIH}
\omega_D=0 \quad \Rightarrow \quad \cZ=\frac{1}{2}{\dot \cU} \;.
\ee
Now we can write the most general invariant action as
\be\label{confaction}
S_{conf}=\kappa \int dt \left( -\omega_K + \alpha m^2 \omega_P +
\beta \nabla_t \Lambda \nabla_t \bLam \omega_P +
i\gamma m \omega_3\right) \;,
\ee
where the covariant derivatives of the fields $\Lambda$ and $\bLam$ are defined as
\be
\left\{ \begin{array}{l} \omega_V=\omega_P \nabla_t \Lambda \\ \bar\omega_V=
\omega_P \nabla_t \bLam \end{array}
\right. \quad \Rightarrow \quad
\left\{ \begin{array}{l} \nabla_t \Lambda = e^{\cU}\frac{ \dot\Lambda}{1+\Lambda\bLam} \\
       \nabla_t \bLam = e^{\cU}\frac{ \dot\bLam}{1+\Lambda\bLam} \end{array} \right.
\ee
The parameters $\alpha, \beta, \gamma$ are dimensionless, while $\kappa^{-1}$ and the constant $m$ have
the dimension of mass $[cm^{-1}]$.
Explicitly, the action \p{confaction} reads
\be\label{confaction1}
S_{conf}=\kappa \int dt \left( \frac{1}{4} e^{\cU}\, {\dot \cU}^2 +
\alpha m^2 e^{-\cU} +
\beta e^{\cU} \frac{\dot\Lambda\dot\bLam}{(1+\Lambda\bLam)^2}+
i\gamma m \frac{\bLam \dot\Lambda- \Lambda \dot\bLam}{1+\Lambda\bLam}  \right) \;.
\ee
This action combines the action of conformal mechanics \cite{AFF}  and that of a charged
particle moving in the
field of a Dirac monopole. With a specific choice of the values of the arbitrary coefficients
$\alpha, \beta$ and $ \gamma$,
the action \p{confaction1} coincides with the bosonic sector of $N=4,8$ superconformal
mechanics \cite{IKL1, BIKL}.
\subsection{Conformal mechanics in the pp-wave limit}
Our strategy is as follows. Firstly, we construct the pp-wave limit for the
$so(1,2)\oplus su(2)$ algebra and
then repeat all steps to find the invariant action.

In order to find the pp-wave limit of the $so(1,2)\oplus su(2)$ algebra \cite{conf1}, we
define the new generators $P_\pm, P_1$,
instead of $P,K,V_3$, as follows:
\be\label{newgen}
P_\pm =\frac{1}{2}\left( P+m^2 K\right) \pm imV_3 \;, \quad P_1=
\frac{1}{2}\left( K- m^{-2} P\right)\;,
\ee
where the constant $m$ has the same dimension $[cm^{-1}]$ as before. We also make
the following rescaling
of all generators:
\be\label{rescaling}
P_+\rightarrow \Omega^2 P_+,\; P_-\rightarrow P_-,\; \left\{ P_1,D,V,\bV\right\} \rightarrow
\left\{\Omega P_1, \Omega D, \Omega V, \Omega \bV\right\} \;.
\ee
The pp-wave limit corresponds to the limit $\Omega \rightarrow 0$ and the algebra
\p{confalg} is reduced, in this limit, to
\bea\label{ppalg}
&& i\left[ P_-, D\right]=m^2 P_1,\quad i\left[ P_1, D\right]=\frac{1}{2m^2}P_+,
\quad i\left[ P_1, P_-\right]=D,\nn
&& i\left[ P_-, V\right]=mV,\quad i\left[ P_-, \bV \right]=-m\bV,\quad i\left[ V,
\bV\right]=-\frac{1}{m}P_+ \;.
\eea

Now, as we did in the previous subsection, we consider a nonlinear realization of
the pp-wave group, with the
algebra \p{ppalg} in its coset over the central charge $P_+$, with an element parameterized as
\be\label{ppcoset}
g=e^{itP_-}e^{i u D}e^{i z P_1 }e^{i\phi V + i\bphi \bV} \;.
\ee
A left shift of the pp-wave group element \p{ppcoset} by
\be
g_2=e^{ia P_-}e^{ib D}e^{ic P_1} e^{if V+i{\bar f}\bV}
\ee
induces the following transformations:
\be\label{pptransf}
\delta t = a,\; \delta u =b \cos (mt) +\frac{c}{m}\sin (mt),\; \delta z =-mb \sin (mt) +
c\cos (mt),\;
\delta \phi = f e^{-imt} \;.
\ee
The left-invariant Cartan forms are given by the following expressions:
\bea\label{ppform}
&& \omega_-=dt\;,\quad \omega_D=du-z dt\;,\quad \omega_1=dz+m^2 u dt\;, \nn
&& \omega_+ = \frac{1}{4}\left( u^2 + \frac{z^2}{m^2}+4\phi\bphi\right)dt -\frac{1}{2m^2}z du +
\frac{i}{2m}\left( \phi d\bphi -\bphi d\phi\right)\; , \nn
&& \omega_V = d\phi+im\, \phi\, dt\;, \bar\omega_V = d\bphi-im\, \bphi\, dt\;.
\eea
Let us note that all coset forms are invariant with respect to \p{pptransf}, while $\omega_+$
is shifted by a full differential
\be
\delta \omega_+ = d\left[ \frac{1}{2m^2}\left( mb \sin (mt)-c \cos (mt)\right) u +
\frac{i}{2m}\left( f e^{-imt}\bphi - {\bar f} e^{imt} \phi \right) \right] \;.
\ee
After expressing the field $z$ in terms of $u$
\be
\omega_D=0 \quad \Rightarrow \quad z= \dot u\;,
\ee
we can write the invariant pp-wave action
\be\label{ppaction1}
S_{pp}= \kappa \int dt \left[ -m^2 \omega_+ + \rho\omega_1 + \sigma\,
\nabla_t \phi \nabla_t \bphi\, \omega_- +
\mu\, m^2\, \omega_- + \nu\, m\,  \omega_V +\bar\nu \,m \,\bar\omega_V
\right],
\ee
where
\be
\nabla_t \phi = \dot\phi+ im\,\phi\;, \quad \nabla_t\bphi = \dot\bphi -im\,\bphi \;.
\ee
Explicitly, the action \p{ppaction1} reads
\bea\label{ppaction2}
S_{pp} &=&  \kappa \int dt \left[ \frac{1}{4} \left( {\dot u}^2 - m^2 u^2\right) -
m^2\phi\bphi +\frac{im}{2}\left(\bphi\dot\phi-\phi\dot\bphi\right) +m^2\rho\, u\right. \nn
&& \left. + \sigma \left( \dot\phi\dot\bphi -im\left( \bphi\dot\phi-
\phi\dot\bphi\right) +m^2\phi\bphi\right)
 + \mu\, m^2 +i\nu\, m^2  \phi -i\bar\nu\, m^2 \bphi \right] \;.
\eea
The action \p{ppaction2} is the most general invariant pp-wave action we could construct.

Before comparing this action
with the conformal invariant action on $AdS_2\times S^2$ \p{confaction1}, we wish to make some
comments.
First of all, the last two terms in \p{ppaction2} cannot be obtained by any reduction procedure from
the action \p{confaction1}. Their
appearance in the pp-wave action \p{ppaction2} becomes admissible, due to the reduction of the
$U(1)$ symmetry generated by
the $V_3$ generator to the central charge $P_+$ in the pp-wave algebra \p{ppalg}. Thus, in order to
compare with the $AdS_2 \times S^2$ action, we will put $\nu=0$.
Secondly, we introduce the new fields $\lambda, \blam$
\be\label{newlambda}
\lambda = e^{imt} \phi, \quad \blam=e^{-imt}\bphi
\ee
and rewrite the action \p{ppaction2} (with $\nu=0$) as
\be\label{ppaction3}
S_{pp} = \kappa \int dt \left[ \frac{1}{4} \left( {\dot u}^2 - m^2 u^2\right) +
\frac{im}{2}\left(\blam\dot\lambda-\lambda\dot\blam\right) +\rho\, m^2 u
 + {\sigma}  \dot\lambda\dot\blam  + \mu\, m^2 \right]\;.
\ee
Let us stress that in the pp-wave action \p{ppaction3} the mass term for the field $u$
and the WZW term come, together with the kinetic term for $u$, from the same Cartan forms
$\omega_-$. As a result, the mass of field $u$ and the coupling constant in front of the
WZW term are completely fixed.

Now, comparing the pp-wave action \p{ppaction3} with the $AdS_2\times S^2$ action \p{confaction1},
one may conclude that all terms in the pp-wave action can be obtained from the
$AdS_2\times S^2$ one by keeping there only terms at most quadratic in the
fields. Additionally, in order to have exact matching, we have to restrict the arbitrary
coefficients, which are present in both actions, as follows:
\be\label{coeff1}
\alpha=-\frac{1}{2},\; \gamma=\frac{1}{2},\; \rho=\frac{1}{2},\; \sigma=\beta,\;
\mu=-\frac{1}{2}.
\ee
Thus, we conclude that the net effect of taking the pp-wave limit in conformal mechanics
consists in reducing all nonlinearities to quadratic ones, together with fixing the value of some
otherwise arbitrary coefficients in the action. Let us note, however, that such a fixing of the coefficients
is not really needed. The actions \p{confaction1} and \p{ppaction3} are invariant with
respect to $so(1,2) \oplus su(2)$ and pp-wave algebra \p{ppalg} transformations
respectively, without any fixing of the coefficients. The relations between coefficients
appear only if we wish to get the pp-wave action from $AdS_2\times S^2$ one.

So, in the pp-wave limit the theories on both sides,
i.e. the "string/particle" and the conformal mechanics side, contain at most quadratic
interaction terms, which represent indeed mass terms for some fields and WZW terms.
In the next section we will demonstrate that the same conclusion is still
valid in the supersymmetric case.

\setcounter{equation}0
\section{N=4 superconformal mechanics in the pp-wave limit}
The construction of $N=4$ superconformal mechanics in the pp-wave limit is similar to the
general consideration of $N=4$ superconformal mechanics, which can be found in \cite{IKL1}.
Precisely, we will start with the superconformal
algebra $su(1,1|2)$
and pass to its pp-wave limit in two steps\footnote{We use the notation of \cite{IKL1}.}
\begin{itemize}
\item Firstly, we redefine the bosonic generator as in \p{newgen} and the spinor generators as
\bea\label{newgenfermi}
&&\wQ^1=\frac{1}{2}\left( Q^1-imS^1\right), \wQ^2=\frac{1}{2}\left( Q^2+imS^2\right), \nn
&& \wS^1=\frac{1}{2m}\left( Q^1+imS^1\right),
\wS^2=\frac{1}{2m}\left( Q^2-imS^2\right)
\eea
\item Secondly, we rescale the bosonic generators as in \p{rescaling} and the spinor ones as
\be\label{rescalingfermi}
\wQ^i \rightarrow \wQ^i \;, \quad \wS^i \rightarrow \Omega \wS^i
\ee
and consider the limit $\Omega \rightarrow 0$.
\end{itemize}
Finally, we get the following pp-wave superalgebra (together with \p{ppalg}):
\bea\label{superalg}
&& \left[P_-, \wS^1\right]=-m\wS^1,\;\left[P_-, \wS^2\right]=m\wS^2,\;
\left[P_1, \wQ^1\right]=-\frac{1}{2}\wS^1,\; \left[P_1, \wQ^2\right]=\frac{1}{2}\wS^2,\nn
&&\left[D, \wQ^i\right]=-\frac{im}{2}\wS^i,\;\left[V, \wQ^1\right]=-im\wS^2,\;
\left[\bV, \wQ^2\right]=im\wS^1, \left\{\wQ^1, \wbS_2\right\}=-i\bV, \nn
&& \left\{\wQ^2, \wbS_1\right\}=-iV,\;\left\{ \wQ^1,\wbS_1\right\}=iD+mP_1,\;
\left\{ \wQ^2,\wbS_2\right\}=-iD+mP_1, \nn
&& \left\{ \wQ^i, \wbQ_j\right\} = -\delta^i_j P_-,\;
\left\{ \wS^i, \wbS_j\right\} = -m^{-2}\delta^i_j P_+.
\eea

Next, we shall construct a nonlinear realization of the pp-wave supergroup with
superalgebra \p{ppalg}, \p{superalg}
on the coset superspace parameterized as
\be\label{ppsupercoset}
g=e^{itP_-}e^{\theta_i\wQ^i+\bar\theta{}^i\wbQ_i}e^{\psi_i\wS^i+
\bpsi^i\wbS_i}e^{i u D}e^{i z P_1 }e^{i\phi V + i\bphi \bV} \;.
\ee
Now, the coordinates $t,\theta_i,\bar\theta{}^i$ parameterize the $N=4,d=1$ superspace,
while the rest of the coset parameters are Goldstone superfields.

The transformation properties of the coordinates and superfields are generated by acting on
the coset element
\p{ppsupercoset} from the left with the elements of the pp-wave group.
The $N=4$ super Poincar\'e transformations are generated by the element
\be\label{g0}
g_0=\exp \left( \varepsilon_1 \wQ^i+ \bar\varepsilon{}^i\wbQ_i \right) \;.
\ee
They read
\be\label{trg0}
\delta t = -\frac{i}{2} \left( \varepsilon_i \bar\theta{}^i+
\bar\varepsilon{}^i\theta_i\right),\;
\delta\theta_i=\varepsilon_i,\; \delta\bar\theta{}^i=\bar\varepsilon{}^i
\ee
and all superfields are scalars.

The transformations under $S$-supersymmetry are generated by the element
\be\label{g1}
g_1=\exp \left( \epsilon_1 \wS^i+ \bar\epsilon{}^i\wbS_i \right)
\ee
and have the following explicit form:
\bea\label{trg1}
&& \delta u=-\bar\theta^1\tilde\epsilon_1+\bar\theta{}^2\tilde\epsilon_2+
\theta_1\overline{\tilde\epsilon}{}^1-
\theta_2\overline{\tilde\epsilon}{}^2,\; \delta z =
-im\left( \bar\theta^i\tilde\epsilon_i+\theta_i\overline{\tilde\epsilon}{}^i\right),\nn
&&\delta\phi=\bar\theta{}^1 \tilde\epsilon_2-\theta_2 \overline{\tilde\epsilon}{}^1,\;
\delta\bar\phi=\bar\theta{}^2 \tilde\epsilon_1-\theta_1 \overline{\tilde\epsilon}{}^2,\;
\delta\psi_1=\tilde\epsilon_1-m\theta_1\theta_2\overline{\tilde\epsilon}{}^2,\;
\delta\psi_2=\tilde\epsilon_2-m\theta_1\theta_2\overline{\tilde\epsilon}{}^1,
\eea
where
\be
\tilde\epsilon_1=\epsilon_1 e^{im{\tilde t}},\;
\tilde\epsilon_2=\epsilon_2 e^{-im{\tilde t}},\;
{\tilde t}=t +\frac{1}{2} \theta_i\bar\theta{}^i\;.
\ee
Since all other super pp-wave group transformations appear in the anticommutators of
Poincar\'{e} and $S$ supersymmetries,
it is sufficient, when constructing the action, to require invariance under these two
supersymmetries.

In what follows we will need the explicit structure of several important Cartan forms,
defined in a standard way
as $g^{-1}d g$
\bea\label{superforms}
&& \omega_- = dt +\frac{i}{2}\left( \theta_i d \bar\theta{}^i +
\bar\theta{}^i d \theta_i \right) \equiv \triangle t , \nn
&& \omega_D =du - z \triangle t - \psi_1 d\bar\theta{}^1 +
\psi_2 d\bar\theta{}^2 +\bpsi{}^1d\theta_1 -\bpsi{}^2 d\theta_2,\nn
&& \omega_V=d\phi +im\phi\triangle t + \psi_2d \bar\theta{}^1 -\bpsi{}^1 d\theta_2,\;
\bar\omega_V=d\bphi -im\bphi\triangle t - \bpsi^2 d \theta_1 +\psi_1 d\bar\theta{}^2.
\eea
Now we may define covariant spinor derivatives as
\be
D^i=\frac{\partial}{\partial\theta_i}+\frac{i}{2}\bar\theta{}^i \partial_t,\;
\bD_i=\frac{\partial}{\partial\bar\theta{}^i}+\frac{i}{2}\theta_i \partial_t,\;
\left\{ D^i,\bD_j\right\}=i \delta^i_j \partial_t \;.
\ee

Prior to constructing the invariant action, one should impose proper irreducibility
constraints on
the $N=4$ superfields. The basic idea for finding the appropriate constraints is the same as in
the case of superconformal
mechanics \cite{IKL1}: we impose constraints such that our basic, low-dimensional
bosonic superfields $u,\phi,\bphi$
contain among their components only four fermions, which should coincide with the
first components
of the spinor superfields $\psi_i,\bpsi{}^i$. These invariant conditions
represent a particular case of the
Inverse Higgs effect \cite{IH} and can be written as
\be\label{superIH}
\omega_D=0\;,\quad \omega_V\left|=0\;,\quad \bar\omega_V\right|=0 \;,
\ee
where $|$ denotes $d\theta$ and $d\bar\theta$ projections of the forms.
Explicitly, the constraints \p{superIH} read
\bea\label{supconstr1}
&& D^1u=-D^2\phi=\bpsi{}^1,\;D^2u=D^1\bphi=-\bpsi{}^2\;,\; D^1\phi=\bD_2\phi=0, \nn
&& \bD_1 u =-\bD_2\bphi = -\psi_1,\;\bD_2 u=\bD_1\phi=\psi_2\;,\; D^2\bphi=\bD_1\bphi=0.
\eea
After introducing a new $N=4$ "vector" superfield $V^{ij}$ via
\be\label{V}
V^{11}=i\sqrt{2}\,\phi,\; V^{22}=-i\sqrt{2}\,\bphi,\;
V^{12}=\frac{i}{\sqrt{2}}u,\quad \overline{V^{ij}}=V_{ij}
\ee
the constraints \p{supconstr1} can be brought in the familiar form
\be\label{supconstr2}
D^{(i}V^{jk)}=0\;, \quad \bD{}^{(i}V^{jk)}=0\;.
\ee
The superfield $V^{ij}$ subject to \p{supconstr2} is recognized as the $N=4, d=1$
tensor multiplet \cite{{mult}, {IKL1}}
with ({\bf 3,4,1}) off-shell components content. Of course, in the present case the
explicit $su(2)$ invariant form
of constraints is a fake, because in the pp-wave limit the $su(2)$ symmetry is
broken down to the algebra
\be
i\left[ V, {\bar V} \right] = -m^{-1}\,P_+
\ee
with $P_+$ being a central charge. Let us note that it is quite natural to have the
same $N=4$ tensor
multiplet as in the case of the $N=4$ superconformal algebra \cite{IKL1}, also in the case of its
pp-wave limit. The reason for this is evident: as we showed in the bosonic case
in the previous Section,
passing to the pp-wave limit means keeping only at most quadratic nonlinearities in
the Lagrangian, provided
that the fields have been chosen properly. Now, in the supersymmetric case we see that
constraints linear in the superfields
are preserved in the pp-wave limit. Thus, it is natural to suggest that the invariant
superfields Lagrangian
should be related with the full $N=4$ superconformal one in the same way as in the
purely bosonic case.
Now we are going to demonstrate that this is really so.

The invariant superfield action consists of a superfield kinetic term and a superpotential.
As in the $N=4$ superconformal mechanics case \cite{IKL1}, the kinetic and potential terms
are easier to write in $N=2$ superspace, let us say the one with coordinates
$\left\{t, \theta_2,\bar\theta{}^2\right\}$.
The analysis of the constraints \p{supconstr1} shows that in the $\theta_1,\bar\theta{}^1$
expansion of the $N=4$
superfields $u,\phi,\bphi$, only the $\theta_1=\bar\theta{}^1=0$ components of each superfield
are independent $N=2$ superfields \cite{IKL1}. Let us denote these superfields as
\be\label{n2sf}
\left. u\right| =v,\; \phi\left|=\rho,\; \bphi\right| = \brho,\quad D^2 \brho=\bD_2 \rho=0 \;,
\ee
where $|$ indicates the $\theta_1=\bar\theta{}^1=0$ restriction.

The transformations of the implicit $N=2$ Poincar\'e supersymmetry generated by
$\wQ_1,\wbQ{}^1$ have the following form, in terms of these $N=2$ superfields:
\be\label{n2tr}
\delta v = \varepsilon{}_1 D^2\rho +
\bar\varepsilon{}^1 \bD_2\brho\;, \; \delta\rho =-\bar\varepsilon{}^1 \bD_2 v\;,\;
\delta\brho=-\varepsilon_1 D^2 v \;,
\ee
while under $\wS$-supersymmetries they transform as
\be\label{n2tr1}
\delta v= -\epsilon_2\bar\theta^2e^{-imt}+\bar\epsilon^2\theta_2 e^{imt},\quad
\delta\rho=\bar\epsilon^1\theta_2 e^{-imt},\;
\delta\bar\rho=-\epsilon_1\bar\theta^2 e^{imt}\;.
\ee

The kinetic and potential terms which are independently invariant with respect
to the implicit $N=2$ Poincar\'{e} supersymmetry \p{n2tr}
can be easily found to be
\bea
&& S_{kin} = A\int dt d^2\theta_2\left( D^2 v \bD_2 v+D^2 \rho\bD_2\bar\rho
\right),  \label{s1} \\
&&S_{pot}= B\int dt d^2\theta_2 \left( v^2- 2\rho\brho\right).\label{s2}
\eea
So, the sum of the actions \p{s1} and \p{s2} possesses $N=4$ super Poincar\'{e}
supersymmetry. Finally, one should check the invariance of the actions with respect
to \p{n2tr1}. After doing so, one can conclude that only the sum of $S_{kin}$ and $S_{pot}$
possesses this symmetry, provided that
\be\label{supercoeff}
2B=mA\;.
\ee
Thus, the invariant action of $N=4$ superconformal mechanics in pp-wave limit
has the following form (with $A=\kappa$):
\be
S_{s-pp} = \kappa \int dt d^2\theta_2\left[ D^2 v \bD_2 v+D^2 \rho\bD_2\bar\rho
+\frac{m}{2} \left( v^2- 2\rho\brho\right)\right]\;,\label{super}
\ee
without {\it any arbitrary coefficients} besides $m$. The invariance with respect to pp-wave
supergroup strictly fixes all possible interacting terms.

It is instructive to rewrite the action \p{super} in terms of components fields,
which can be defined as
\be\label{comp}
v|,\; \psi=iD^2 v|,\;\bpsi=i\bD_2 v|,\; {\cal A}=\left[D^2,\bD_2\right]v|,\quad
\xi=D^2\rho|,\; \bxi=\bD_2\rho|\;,
\ee
where $|$ means $\theta_2=\bar\theta^2=0$. Integrating over $\theta$ in \p{super}
and eliminating the auxiliary field $\cal A$ by making use of its equation of motion, we end up
with the following action:
\be\label{compaction}
S_{s-pp}=\kappa \int dt\left[\frac{1}{4}{\dot v}^2 -\frac{m^2}{4}v^2+
{\dot\rho}{\dot{\bar\rho}}+\frac{im}{2}\left(\rho\dot{\bar\rho}-{\dot\rho}\bar\rho
\right)
 +i\dot\psi\bpsi +i\dot\xi\bxi +m\left( \psi\bpsi-\xi\bxi\right)\right].
\ee

Finally, one can conclude that in the fully supersymmetric case the pp-wave action
represents a limiting case of the $N=4$ superconformal
mechanics action, where only nonlinearities at most quadratic in the superfields survive.

\setcounter{equation}0
\section*{Conclusion}
Motivated by the interest in understanding the role of AdS/CFT duality in the
Penrose limit, whose "conformal" side has received so far much less attention
than the corresponding "string" side,
we constructed in this paper the pp-wave limit of $N=4$ superconformal mechanics with the
off-shell $({\bf 3,4,1})$ multiplet. We showed that this multiplet can be described
by a properly constrained Goldstone superfield, associated with a suitable coset of the
nonlinearly realized $N=4,d=1$ pp-wave supergroup. We presented the superfield and
the component actions, which exhibit the interesting property that the interaction
part is completely fixed by symmetry.
Moreover, for the pp-wave case, the kinetic and potential terms are invariant
only when taken together, as a linear combination of the two terms, provided
the value if their relative coefficient is appropriately set.
We also explicitly demonstrated that the passing to pp-wave limit can be achieved by
keeping at most quadratic nonlinearities in the action of (super)conformal mechanics.

\section*{Acknowledgments}
This work was partially supported by the European
Community's Human Potential
Programme under contract HPRN-CT-2000-00131 Quantum Spacetime,
the INTAS-00-00254 grant, the NATO Collaborative Linkage Grant PST.CLG.979389,
RFBR-DFG grant No 02-02-04002, grant DFG No 436 RUS 113/669, RFBR grant
No 03-02-17440 and a grant of the Heisenberg-Landau programme.

\end{document}